\documentclass{iau}

\usepackage[english]{babel}
\usepackage[latin1]{inputenc}
\usepackage{amssymb}
\usepackage{graphicx}
\usepackage{color}
\usepackage{latexsym}
\usepackage{multirow}
\usepackage{txfonts}
\usepackage{amstext}

\def\lsim{\mathrel{\rlap{\lower3.5pt\hbox{\hskip0.5pt$\sim$}}     \raise0.5pt\hbox{$<$}}}                

\def\zs{\mbox{{$z_{\rm spec}$}}}
\def\zp{\mbox{{$z_{\rm phot}$}}}

\title[Cooperative photo-z estimation]{Cooperative photometric redshift estimation}
\author[Cavuoti S. et al. ]{S.~Cavuoti$^{1}$, C.~Tortora$^{2}$, M.~Brescia$^{1}$, G.~Longo$^{3}$, M.~Radovich$^{4}$, N.~R.~Napolitano$^{1}$, V. Amaro$^{3}$, \and  C. Vellucci$^{5}$}
\affiliation{
$^{1}$INAF - Astronomical Observatory of Capodimonte, via Moiariello 16, I-80131 Napoli, Italy\\[\affilskip]
$^{2}$Kapteyn Astronomical Institute, Univ. of Groningen, P.O. Box 800, 9700 AV Groningen, the Netherlands\\[\affilskip]
$^{3}$Department of Physics, University Federico II, Via Cinthia 6, I-80126 Napoli, Italy\\[\affilskip]
$^{4}$INAF - Astronomical Observatory of Padua, vicolo dell'Osservatorio 5, I-35122 Padova, Italy\\[\affilskip]
$^{5}$ DIETI, University of Naples Federico II, Via Claudio,21 I-80125 Napoli, Italy}

\pubyear{2017}
\volume{325}  
\jname{Astroinformatics 2016}
\editors{M. Brescia, S.G. Djorgovski, E. Feigelson, \\ G. Longo, \& S. Cavuoti,  eds.}
\begin{document}

\maketitle

\begin{abstract}
In the modern galaxy surveys photometric redshifts play a central role in a broad range of studies, from gravitational lensing and dark matter distribution to galaxy evolution.
Using a dataset of $\sim25,000$ galaxies from the second data release of the Kilo Degree Survey (KiDS) we obtain photometric redshifts with five different methods: \textit{(i)} Random forest, \textit{(ii)} Multi Layer Perceptron with Quasi Newton Algorithm, \textit{(iii)} Multi Layer Perceptron with an optimization network based on the Levenberg-Marquardt learning rule, \textit{(iv)} the Bayesian Photometric Redshift model (or BPZ) and \textit{(v)} a classical SED template fitting procedure (Le Phare). We show how SED fitting techniques could provide useful information on the galaxy spectral type which can be used to improve the capability of machine learning methods constraining systematic errors and reduce the occurrence of catastrophic outliers. We use such classification to train specialized regression estimators, by demonstrating that such hybrid approach, involving SED fitting and machine learning in a single collaborative framework, is capable to improve the overall prediction accuracy of photometric redshifts.

\end{abstract}

\firstsection
\section{Introduction}
Photometric redshift produced through  the modern multi-band digital sky surveys are crucial to provide a reliable distance estimation for a large number of galaxies in order to be used for several tasks in precision cosmology, to mention just a few: the weak gravitational lensing to constrain dark matter and dark energy, the identification of galaxy clusters and groups, the search of strong lensing and ultra-compact galaxies, as well as the study of the mass function of galaxy clusters.
We can derive photometric redshift (hereafter photo-z) thanks to the existence of a hidden (and complex) correlation among the fluxes in the different broad bands, the spectral types of the object itself, and the real distance.
Although hidden, the mapping function that can map the photometric space into the redshift one could be approximated in several ways, and the existing methods can be broadly divided into two main classes: theoretical and empirical.

In the previous work \cite[(Cavuoti et al. 2015a)]{Cavuoti+15_KIDS_I} we have already applied an empirical method, the Multi Layer Perceptron with Quasi Newton Algorithm, MLPQNA, (\cite[Cavuoti et al. 2012]{cavuoti2012}, \cite[Brescia et al. 2013]{brescia2013}, \cite[Brescia et al. 2014]{brescia2014},
\cite[Brescia et al. 2015]{brescia2015}, \cite[Cavuoti et al. 2017]{cavuoti2017}), to a dataset extracted from the Kilo Degree Survey (KiDS).
Here we apply five different photo-z techniques to the same dataset and then we analyze the behavior of such methods with the aim at finding a way to combine their features in order to to optimize the accuracy of photo-z estimation; a similar, but reversed approach was followed recently by \cite{fotopoulou2016}.

\section{The data}

As stated before we used the photometric data from the KiDS optical survey \cite[(de Jong et al. 2015)]{deJong+15_KIDS_paperI}. The KiDS data releases consist of tiles which are observed in the \textit{u, g, r}, and \textit{i} bands.
The sample of galaxies on which we performed our analysis is mostly extracted from KiDS-DR2 \cite[(de Jong et al. 2015)]{deJong+15_KIDS_paperI}, which contains $148$ tiles observed in all filters during the first two years of survey regular operations. We added $29$ extra tiles, not included in the DR2 at the time this was released, that will be part of the forthcoming KiDS data release, thus covering an area of $177$ square degrees.

We used the multi-band source catalogs, based on source detection in the \textit{r}-band images. While magnitudes are measured in all filters, the star-galaxy separation, as well as the positional and shape parameters are derived from the r-band data only, which typically offers the best image quality and seeing $\sim 0.65''$, thus providing the most reliable source positions and shapes. The KiDS survey area is split into two fields, KiDS-North and KiDS-South, KiDS-North is completely covered by the combination of SDSS and the 2dF Galaxy Redshift Survey (2dFGRS), while KiDS-South corresponds to the 2dFGRS south Galactic cap region. Further details about data reduction steps and catalog extraction are provided in \cite{deJong+15_KIDS_paperI} and \cite{Tortora+15_KiDS_compacts}.

Aperture photometry in the four \textit{ugri} bands measured within several radii was derived using S-Extractor \cite[(Bertin \& Arnouts 1996)]{bertin1996}. In this work we use magnitudes ${\tt MAGAP\_4}$ and ${\tt MAGAP\_6}$, measured within the apertures of diameters $4''$ and $6''$, respectively. These apertures were selected to reduce the effects of seeing and to minimize the contamination from mis-matched sources. The limiting magnitudes are: MAGAP\_4\_u $=25.17 $,  MAGAP\_6\_u $=24.74 $,   MAGAP\_4\_g $=26.03 $,   MAGAP\_6\_g $=25.61 $,   MAGAP\_4\_r $=25.89 $,   MAGAP\_6\_r $=25.44 $,   MAGAP\_4\_i $=24.53 $,   MAGAP\_6\_i $=24.06 $. To correct for residual offsets in the photometric zero points, we used the SDSS as reference: for each KiDS tile and band we matched bright stars with the SDSS catalog and computed the median difference between KiDS and SDSS magnitudes (\textit{psfMag}). For more details about data preparation and pre-processing see \cite{deJong+15_KIDS_paperI} and \cite{Cavuoti+15_KIDS_I}.

In order to build the spectroscopic Knowledge Base (KB) we cross-matched the KiDS data with the spectroscopic samples available in the GAMA data release $2$, \cite[(Liske et al. 2015)]{liske2015}, and SDSS-III data release $9$ \cite[(Ahn et al. 2012)]{ahn2012}.
The detailed procedure adopted to obtain  the data  used for the experiments was as follow: \textit{(i)} we excluded objects having low photometric quality (i.e., with flux error higher than one magnitude); \textit{(ii)} we removed all objects having at least one missing band (or labeled as Not-a-Number or NaN), thus obtaining the cleaned catalogue used to create the training and test sets, in which all photometric and spectroscopic information required is complete for all objects; \textit{(iii)} we performed a randomly shuffled splitting into a training and a blind test set, by using the $60\% / 40\%$ percentages, respectively; \textit{(iv)} we applied the cuts on limiting magnitudes (see \cite[Cavuoti et al. 2015b]{cavuoti2015} for details): \textit{(v)} we selected objects with {\tt IMA\_FLAGS} equal to zero in the \textit{g, r} and \textit{i} bands, i.e., sources that have not been flagged because located in proximity of saturated pixels, star haloes, image border or reflections, or within noisy areas, see \cite{deJong+15_KIDS_paperI}. The \textit{u} band is not considered in such selection since the masked regions relative to this band are less extended than in the other three KiDS bands.

The final KB consists of $15,180$ objects to be used as training set and $10,067$ for the test set.

\section{The methods}

We chose three machine learning methods, among the ones which are publicly available in the DAta Mining \& Exploration Web Application REsource or simply DAMEWARE \cite[(Brescia et al. 2014)]{brescia2014} web-based infrastructure: the Random Forest (RF; \cite[Breiman 2001]{breiman2001}), and two versions of the Multi Layer Perceptron with different optimization methods, i.e., the Quasi Newton Algorithm \cite[(Byrd et al. 1994)]{byrd1994} and the Levenberg-Marquardt rule \cite[(Nocedal \& Wright 2006)]{nocedal2006}, respectively; furthermore we made use of a SED fitting method: Le Phare \cite[(Ilbert et al. 2006)]{ilbert2006} and BPZ \cite[(Benitez 2000)]{Benitez}, a Bayesian photo-z estimation based on a template fitting method which is the last method involved in our experiments.
The results were evaluated using only the objects of the blind test set by calculating the following set of standard statistical estimators for the quantity $\Delta z = (\zs-\zp)/(1+\zs)$: \textit{(i)} bias: defined as the mean value of the residuals $\Delta z$;
\textit{(ii)} $\sigma$: the standard deviation of the residuals; \textit{(iii)} $\sigma_{68}$: the radius of the region that includes $68\%$ of the residuals close to 0; \textit{(iv)} NMAD: Normalized Median Absolute Deviation of the residuals, defined as $NMAD(\Delta z) = 1.48 \times Median (|\Delta z|)$;
\textit{(v)} fraction of outliers with $|\Delta z| > 0.15$.

\section{Experiments}
\begin{figure}
\centering
\includegraphics[width=0.71\textwidth]{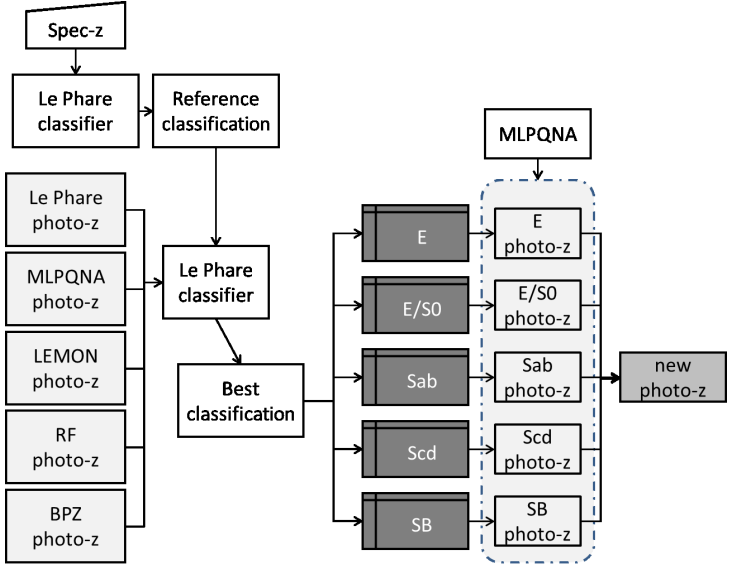}
\caption{Workflow of the method implemented to combine SED fitting and ML models to improve the overall photo-z estimation quality. See text for details.}\label{fig:workflow}
\end{figure}

After a preliminary evaluation of the photometric redshifts, based on each of the five methods, by analyzing the results on the basis of the spectral type classification performed by Le Phare (i.e., the class of the template which shows the best fitting), we noticed that ML methods have a better performance, although strongly dependent from the spectral type itself.
Therefore, we decide to exploit the capability of Le Phare to produce such spectral type classifications to train a specific regressor for each class. The workflow is described in Fig.~\ref{fig:workflow}.
It goes without saying that the training of a specific regression model for each class can be effective only if the subdivision itself is as accurate as possible.

After having obtained the preliminary results, we started by creating a reference spectral type classification of data objects through Le Phare model. By bounding the fitting procedure with the spec-z's, Le Phare provided the templates with the best fit. In this way it was possible to assign a specific spectral type class to each object. Afterwards, by replacing the spec-z's with the photo-z's estimated by the preliminary experiment and by alternating the $5$ photo-z estimates (one for each applied model) as redshift constraint for the fitting procedure, the Le Phare model was used to derive five different spectral type classifications for each object of the KB.

A normalized confusion matrix has been used to find the best classification, as the class of the best fit template derived from Le Phare which resulted from the experiment using the photometric redshifts produced by the random forest.
By comparing the five matrices, the case of RF model 
presents the best behavior for all classes. Therefore, we considered as the best classification the one obtained by using the photo-z's provided by the RF model.
We then subdivided the KB on the base of the five spectral type classes, thus obtaining five different subsets used to perform distinct training and blind test experiments, one for each individual class. 

The final stage of the workflow consisted into the combination of the five subsets to produce the overall photo-z estimation, which was compared with the preliminary experiment in terms of the statistical estimators described in Sec. 3. The combined statistics was calculated on the whole datasets, after having gathered together all the objects of all classes and are reported in the last two rows of the Table~\ref{tab:class}. As with single classes, all the statistical estimators show an improvement in the combined approach case, with the exception of a slightly worst bias.
As Table~\ref{tab:class} shows, the proposed combined approach induces an estimation improvement for each class, as well as for the whole dataset.

\begin{table}
\centering
\begin{tabular}{lrrrrrrr}
\hline
  \textbf{$Class$} & \textbf{$Exptype$} & 	  \textbf{$Datasize$} &  \textbf{$bias$} &  \textbf{$\sigma$} &  \textbf{$NMAD$} &   \textbf{$out. (\%)$} & \textbf{$\sigma68$} \\
\hline
  \textbf{E} & \textbf{hybrid} 	& \textbf{638} 	& \textbf{-0.0009} & \textbf{0.020} & \textbf{0.016} & \textbf{0.00} & \textbf{0.017}\\
  E & standard 	& 638 	&  0.0130 & 0.029 & 0.022 & 0.31 & 0.028\\
  \textbf{E/S0} & \textbf{hybrid}	& \textbf{2858} 	& \textbf{-0.0005} & \textbf{0.016} & \textbf{0.012} & \textbf{0.10} & \textbf{0.012}\\
  E/S0 & standard 	& 2858 	& -0.0059 & 0.022 & 0.014 & 0.31 & 0.014\\
  \textbf{Sab} & \textbf{hybrid}	& \textbf{1383} 	& \textbf{-0.0003} & \textbf{0.015} & \textbf{0.015} & \textbf{0.00} & \textbf{0.014}\\
  Sab & standard 	& 1383 	& -0.0032 & 0.018 & 0.016 & 0.00 & 0.016\\
  \textbf{Scd} & \textbf{hybrid} 	& \textbf{3900} 	& \textbf{-0.0011} & \textbf{0.024} & \textbf{0.019} & \textbf{0.18} & \textbf{0.019}\\
  Scd & standard 	& 3900 	&  0.0006 & 0.025 & 0.020 & 0.23 & 0.020\\
  \textbf{SB} & \textbf{hybrid} 	& \textbf{1288} 	& \textbf{-0.0014} & \textbf{0.038} & \textbf{0.021} & \textbf{0.70} & \textbf{0.022}\\
  SB & standard 	& 1288 	&  0.0027 & 0.038 & 0.022 & 0.85 & 0.023\\\hline
  \textbf{ALL} & \textbf{hybrid} 	& \textbf{10067}	& \textbf{-0.0008} & \textbf{0.023} & \textbf{0.016} & \textbf{0.19} & \textbf{0.016}\\
  ALL & standard 	& 10067 & -0.0007 & 0.026 & 0.018 & 0.31 & 0.018\\
\hline
\end{tabular}
\caption{Photo-z estimation results based on MLPQNA model for each spectral type subset of the test set, classified by Le Phare by bounding the fit through the photo-z's predicted by RF model, which provided the best classification. The term \textit{hybrid} refers to the results obtained by the workflow discussed here and based on the combined approach, while \textit{standard} refers to the results obtained on the same objects but through the standard approach.}\label{tab:class}
\end{table}

\section{Discussion and conclusions}

In this work we described an original workflow designed to improve the photo-z estimation accuracy through a combined use of theoretical (SED fitting) and empirical (machine learning) methods.
The data sample used for the analysis was extracted from the ESO KiDS DR2 photometric galaxy data, using a knowledge base derived from the SDSS and GAMA spectroscopic samples. For a catalog of about $25,000$ galaxies with spectroscopic redshifts, we estimated photo-z's using five different methods: \textit{(i)} Random Forest; \textit{(ii)} MLPQNA (Multi Layer Perceptron with the Quasi Newton learning rule); \textit{(iii)} LEMON (Multi Layer Perceptron with the Levenberg-Marquardt learning rule); \textit{(iv)} Le Phare SED fitting and \textit{(v)} the bayesian model BPZ. The results obtained with the MLPQNA on the complete KiDS DR2 data have been discussed in \cite{Cavuoti+15_KIDS_I}, and further details are provided there.

The spectral type classification provided by the SED fitting method allows to derive also for ML models the statistical errors as function of spectral type, thus leading to a more accurate characterization of the errors. Therefore, it is possible to assign a specific spectral type attribute to each object and to evaluate single class statistics. This fact by itself, can be used to derive a better characterization of the errors. Furthermore, as it has been shown, the combination of SED fitting and ML methods allows also to build specialized (i.e., expert) regression models for each spectral type class, thus refining the process of redshift estimation.

Although the spec-z's are in principle the most accurate information available to bound the SED fitting techniques, this would make impossible to produce a wide catalogue of photometric redshifts, that would also include objects not observed spectroscopically. Thus, it appears reasonable to identify the best solution by making use of predicted photo-z's to bound fitting, in order to obtain a reliable spectral type classification for the widest set of objects. This approach, having also the capability to use arbitrary ML and SED fitting methods, makes the proposed workflow widely usable in any survey project.

By looking at Table~\ref{tab:class}, our procedure shows clearly how the MLPQNA regression method benefits from the knowledge contribution provided by the combination of SED fitting (Le Phare in this case) and machine learning (RF in the best case. This allows to use a set of regression experts based on MLPQNA model, specialized to predict redshifts for objects belonging to specific spectral type classes, thus gaining in terms of a better photo-z estimation.

By analyzing the results of Table~\ref{tab:class} in more detail, the improvement in photo-z quality is significant for all classes and for all statistical estimators. Only the two classes \textit{Scd} and \textit{SB} show a less evident improvement, since their residual distributions appear almost comparable in both experiment types, as confirmed by their very similar values of statistical parameters $\sigma$ and $\sigma_{68}$.
This leads to obtain a more accurate photo-z prediction by considering the whole test set.

The only apparent exception is the mean (column \textit{bias} of Table~\ref{tab:class}), which suffers the effect of the alternation of positive and negative values in the \textit{hybrid} case, that causes the algebraic sum to result slightly worse than the \textit{standard} case (the effect occurs on the fourth decimal digit, see column \textit{bias} of the last two rows of Table~\ref{tab:class}). This is not statistically relevant because the bias is one order of magnitude smaller than $\sigma$ and $\sigma_{68}$, therefore negligible.

We note that in some cases, the \textit{hybrid} approach leads to the almost complete disappearance of catastrophic outliers. This is the case, for instance of the \textit{E} type galaxies. The reason is that for the elliptical galaxies the initial number of objects is lower than for the other spectral types in the KB. In the \textit{standard} case, i.e., the standard training/test of the whole dataset, such small amount of \textit{E} type representatives is mixed together with other more populated class objects, thus causing a lower capability of the method to learn their photometric/spectroscopic correlations. Instead, in the \textit{hybrid} case, using the proposed workflow, the possibility to learn \textit{E} type correlations through a regression expert increases the learning capabilities, thus improving the training performance and the resulting photo-z prediction accuracy.

The confusion matrices allow us to compare classification statistics.  The most important statistical estimators are: \textit{(i)} the \textit{purity} or \textit{precision}, defined as the ratio between the number of correctly classified objects of a class (the block on the main diagonal for that class) and the number of objects predicted in that class (the sum of all blocks of the column for that class); \textit{(ii)} the \textit{completeness} or \textit{recall}, defined as the ratio between the number of correctly classified objects in that class (the block on the main diagonal for that class) and the total number of (true) objects of that class originally present in the dataset (the sum of all blocks of the row for that class); \textit{(iii)} the \textit{contamination}, automatically defined as the reciprocal value of the \textit{purity}.

\textit{Scd} and \textit{SB} spectral type classes are well classified by all methods. This is also confirmed by their statistics, since the \textit{purity} is on average on all five cases around $88\%$ for \textit{Scd} and $87\%$ for \textit{SB}, with an averaged \textit{completeness} of, respectively, $91\%$ in the case of \textit{Scd} and $82\%$ for \textit{SB}.

Moreover the three classifications based on the machine learning models maintain a good performance in the case of \textit{E/S0} spectral type class, reaching on average a \textit{purity} and a \textit{completeness} of $89\%$ for both estimators.

In the case of \textit{Sab} class, only the RF-based classification is able to reach a sufficient degree of efficiency ($78\%$ of \textit{purity} and $85\%$ of \textit{completeness}). In particular, for the two cases based on photo-z's predicted by SED fitting models, for the \textit{Sab} class the BPZ-based results are slightly more \textit{pure} than those based on Le Phare ($68\%$ vs $66\%$) but much less \textit{complete} (49\% vs 63\%).

Finally, by analyzing the results on the \textit{E} spectral type class, the classification performance is on average the worst case, since only the RF-based case is able to maintain a sufficient compromise between \textit{purity} ($77\%$) and \textit{completeness} ($63\%$). The classification based on Le Phare photo-z's reaches a $69\%$ of completeness on the \textit{E} class, but shows an evident high level of contamination between \textit{E} and \textit{E/S0}, thus reducing its purity to the $19\%$. We also note that the intrinsic major difficulty to separate \textit{E} objects from \textit{E/S0} class is due to the partial co-presence of both spectral types in the class \textit{E/S0}, that may partially cause wrong evaluations by the classifier.

Furthermore, the fact that later Hubble types are less affected may be easily explained by considering that their templates are, on average, more homogeneous than for early type objects.

All the above considerations lead to the clear conclusion that the classification performed by Le Phare model and based on RF photo-z's achieves the best compromise between purity and completeness of all spectral type classes. Therefore, its spectral classification has been taken as reference throughout the further steps of the workflow.

At the final stage of the proposed workflow, the photo-z quality improvements obtained by the expert MLPQNA regression estimators on single spectral types of objects induce a reduction of $\sigma$ from $0.026$ to $0.023$ and of $\sigma_{68}$ from $0.018$ to $0.016$ for the overall test set, 
in addition to a more significant improvement for the E class ($\sigma$ from $0.029$ to $0.020$ and of $\sigma_{68}$ from $0.028$ to $0.017$). This is mostly due to the reduction of catastrophic outliers. This  result, together with the generality of the workflow in terms of choice of the classification/regression methods, demonstrates the possibility to optimize the accuracy of photo-z estimation through the collaborative combination of theoretical and empirical methods.

\section*{Acknowledgments}
CT is supported through an NWO-VICI grant (project number $639.043.308$). MB and SC acknowledge financial contribution from the agreement ASI/INAF I/023/12/1. MB acknowledges the PRIN-INAF 2014 {\it Glittering kaleidoscopes in the sky: the multifaceted nature and role of Galaxy Clusters}.

\end{document}